\documentclass[aps,prl,reprint,amsmath,amssymb,superscriptaddress]{revtex4-1}

\usepackage{graphicx}
\usepackage{dcolumn}
\usepackage{bm}
\usepackage{xcolor}

\usepackage[mathscr]{euscript}

\DeclareSymbolFont{rsfs}{U}{rsfs}{m}{n}
\DeclareSymbolFontAlphabet{\mathscrsfs}{rsfs}

\begin{document}

\title{Tuning the Effective $\mathcal{PT}$-Phase of Plasmonic Eigenmodes}

\author{Alessandro Tuniz}
\affiliation{Institute of Photonics and Optical Science (IPOS) and the University of Sydney Nano Institute (Sydney Nano), School of Physics, University of Sydney, NSW 2006, Australia}
\email[]{alessandro.tuniz@sydney.edu.au}
\author{Torsten Wieduwilt}
\affiliation{Leibniz Institute of Photonic Technology (IPHT Jena), Albert-Einstein-Str. 9, 07745 Jena, Germany}
\author{Markus A. Schmidt}
\affiliation{Leibniz Institute of Photonic Technology (IPHT Jena), Albert-Einstein-Str. 9, 07745 Jena, Germany}

\date{\today}

\begin{abstract}
We experimentally observe an effective $\mathcal{PT}$-phase transition through the exceptional point in a hybrid plasmonic-dielectric waveguide system. Transmission experiments reveal fundamental changes in the underlying Eigenmode interactions as the environmental refractive index is tuned, which can be unambiguously attributed to a crossing through the plasmonic exceptional point. These results extend the design opportunities for tuneable non-Hermitian physics to plasmonic systems.
\end{abstract}

\maketitle

Hermitian systems and their operators are used to describe a wide range of physical phenomena to predict the evolution of Eigenstates via unitary operations containing purely real Eigenvalues~\cite{dunford1957linear, shankar2012principles}. Recently, increasing attention has been dedicated to non-Hermitian systems, which are non-conservative and generally yield complex Eigenvalue spectra~\cite{heiss2012physics}.
Non-Hermitian systems are typically created by opening a Hermitian system to the environment by including dissipation and/or gain. Interestingly, coupled non-Hermitian systems can yield purely real Eigenvalues which respect parity-time ($\mathcal{PT}$) symmetry via an appropriate balance of gain and loss~\cite{bender1998real, feng2017non}. In an unbalanced scenario, a $\mathcal{PT}$-broken system shows complex conjugate Eigenvalue pairs. The $\mathcal{PT}$-symmetric (PTS) and $\mathcal{PT}$-broken (PTB) regimes are separated by the \emph{exceptional point} (EP), where the Eigenvalues coalesce,
which is associated with several interesting physical phenomena, such as level repulsion~\cite{heiss2000repulsion}, reflectionless propagation~\cite{feng2013experimental}, and topological phase transitions~\cite{rudner2009topological}.

Photonics has been identified as the ideal landscape for investigating the subtle properties of non-Hermitian systems~\cite{feng2017non,el2018non,miri2019exceptional,ozdemir2019parity}, since tailored amounts of optical gain and loss can be introduced by appropriately engineered materials and structures. Many important technological advances have recently been provided by EP-photonics, such as sensitive modal manipulation~\cite{hodaei2014parity}, topological energy transfer~\cite{xu2016topological}, unidirectional propagation~\cite{wang2009observation}, and polarization conversion~\cite{hassan2017dynamically}. These advances rely on the unique characteristics of the Eigenstates close to the EP -- for example, even nanoscale events can be detected by measuring their impact on macroscopic Eigenstates~\cite{chen2017exceptional}.

For many applications, balancing gain and loss is not necessary for accessing the underlying physics. Most investigations~\cite{guo2009observation, feng2013experimental,zhen2015spawning,feng2017non} harness \emph{effective} $\mathcal{PT}$-symmetric systems with a global loss offset, where PTS and PTB states, separated by an EP, are achieved with Eigenvalues that are shifted along the imaginary axis with respect to the perfectly balanced case. Such effective $\mathcal{PT}$-symmetric structures open up many opportunities in the design of photonic systems that harness $\mathcal{PT}$-symmetries without requiring gain, simplifying designs.

One state that can be used for non-Hermitian photonics is the surface plasmon polariton (SPP)~\cite{nov}, which is a propagating surface wave at metal/dielectric interface. SPPs continue to attract attention due to their ability to confine light down to a fraction of the wavelength, and their high sensitivity to RI changes of the environment~\cite{nov,wieduwilt2015ultrathin}. Owing to their large wave-vectors, coupling energy to SPPs is often challenging. One approach relies on using (lossless) dielectric waveguides running parallel to (lossy) plasmonic waveguide~\cite{tuniz2016broadband}, leading to a system with hybrid Eigenmodes (EMs) and complex Eigenvalues. Such dielectric/plasmonic systems are inherently non-Hermitian, and can be designed to possess an effective-$\mathcal{PT}$ phase transition -- full $\mathcal{PT}$-symmetric properties can then be recovered by re-introducing gain~\cite{alaeian2014non,salgueiro2016optimization}.
Despite its great potential in a multitude of areas~\cite{miri2019exceptional}, and a growing number of studies~\cite{alaeian2014non,salgueiro2016optimization,kodigala2016exceptional,turitsyna2017guided,lourencco2018self}, to the best of our knowledge effective $\mathcal{PT}$-phase transitions, tuned across the EP, have yet to be experimentally reported in plasmonic waveguides.
Here, we experimentally observe the effective $\mathcal{PT}$-phase transition across the plasmonic EP of a tuneable hybrid dielectric/plasmonic waveguide system. Transmission experiments reveal fundamental changes in the properties of the underlying system by tailoring the refractive index of the surrounding environment and unambiguously showing a crossing through the plasmonic EP.

\begin{figure}[t!]
\includegraphics[width=\linewidth]{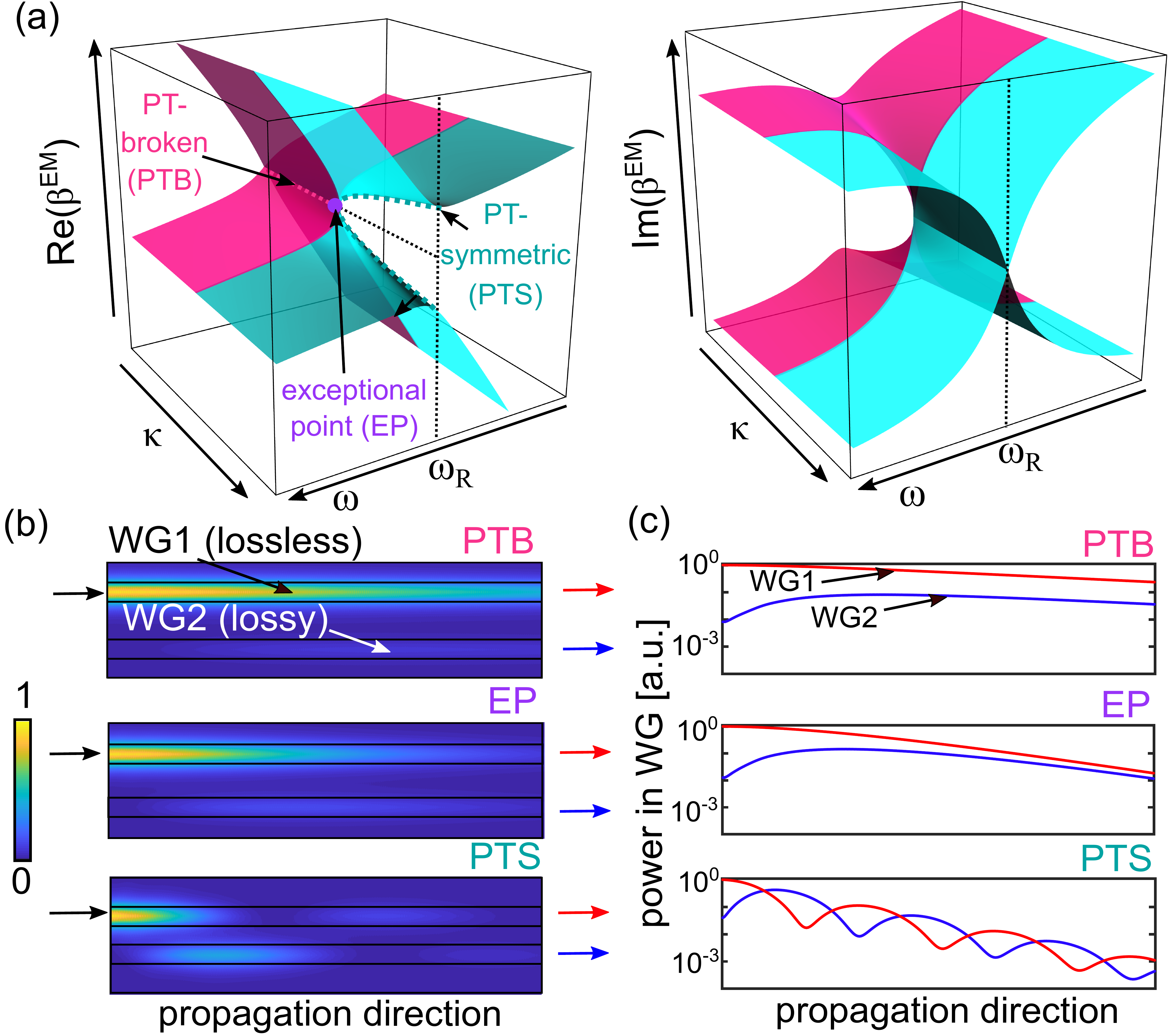}
\caption{\label{fig:fig1}(a) Real and imaginary part of $\beta_j^{\rm EM}$ calculated with the CMT model. (b) Example FEM calculations of the coupling behaviour between a lossless and lossy WG at the phase matching point, highlighting the PTB and PTS regimes separated by the EP  (black arrow: WG1 input; red and blue arrows: WG1 and WG2 output, respectively). 
(c) Axial power distribution in each waveguide, in the three regimes considered in (b). See SM for further details.}
\end{figure}

The main features of such system can be understood on the basis of mode hybridization using coupled mode theory (CMT). We consider the specific case of a lossless waveguide 1 (WG1) and lossy WG2, supporting modes with respective propagation constants $\beta_1 = \beta_{1}^R$ and  $\beta_2 = \beta_{2}^R + i\beta_{2}^I$, and coupling constant $\kappa$, which for this particular case is assumed to be real~\cite{miri2019exceptional}. The coupled mode equations can be written as

\begin{eqnarray}
\frac{d}{dz}
\begin{pmatrix}
a_1\\
a_2
\end{pmatrix}
= -i
\begin{pmatrix}
\beta_1^R & \kappa \\
\kappa & \beta_2^R + i\beta_2^I
\end{pmatrix}
\begin{pmatrix}
a_1\\
a_2
\end{pmatrix}
\label{eq:one}
\end{eqnarray}
where $a_{1,2}$ are the uncoupled modal field amplitudes in each waveguide. Solutions of the form $a_j = A_j \, \exp(i \beta^{\rm EM}_j z)$ lead to hybrid EMs with propagation constants

\begin{eqnarray}
\beta_j^{\rm EM} = \overline{\beta_R} + i \frac{\beta_2^I}{2} \pm \sqrt{\kappa^2 + \left(\Delta \beta_R + i \frac{\beta_2^I}{2} \right)^2},
\end{eqnarray}
where $\overline{\beta_R} =  \left(\beta_1^R +  \beta_2^R\right)/2$, $\Delta \beta_R =  \left(\beta_1^R -  \beta_2^R\right)/2$ is the dephasing, and $j=1,2$. We find that the CMT model is in excellent agreement with simulations of a lossy 1D multilayer system~\cite{tuniz2016broadband} (see SM) 
The  model predicts different transmission characteristics near the EP, which occurs when the Eigenvalues coalesce at a certain frequency $\omega_R$ such that $\sqrt{\kappa^2 + \left(\Delta \beta_R + i \beta_2^I/2 \right)^2}=0$, so that $\Delta{\beta_R} = 0$ (i.e., the phase-matching condition) and $\kappa = \beta_2^I/2$. Figure~\ref{fig:fig1}(b)-(c) shows finite element (FE) calculations which highlight the transmission characteristics close to the EP: the PTS regime is characterized by a periodic exchange of energy between the waveguides due to modal beating between EMs with different propagation constants -- analagously to what occurs in Hermitian systems. The PTB regime is unique to non-Hermitian systems and presents no periodic energy exchange between waveguides at $\omega_R$, since the real parts of the coupled Eigenmodes are equal, resulting in an infinite beat length. These different regimes can be accessed at $\omega_R$ by changing $\kappa/\beta_2^I$.

\begin{figure}[b!]
\centering
\includegraphics[width=0.9\linewidth]{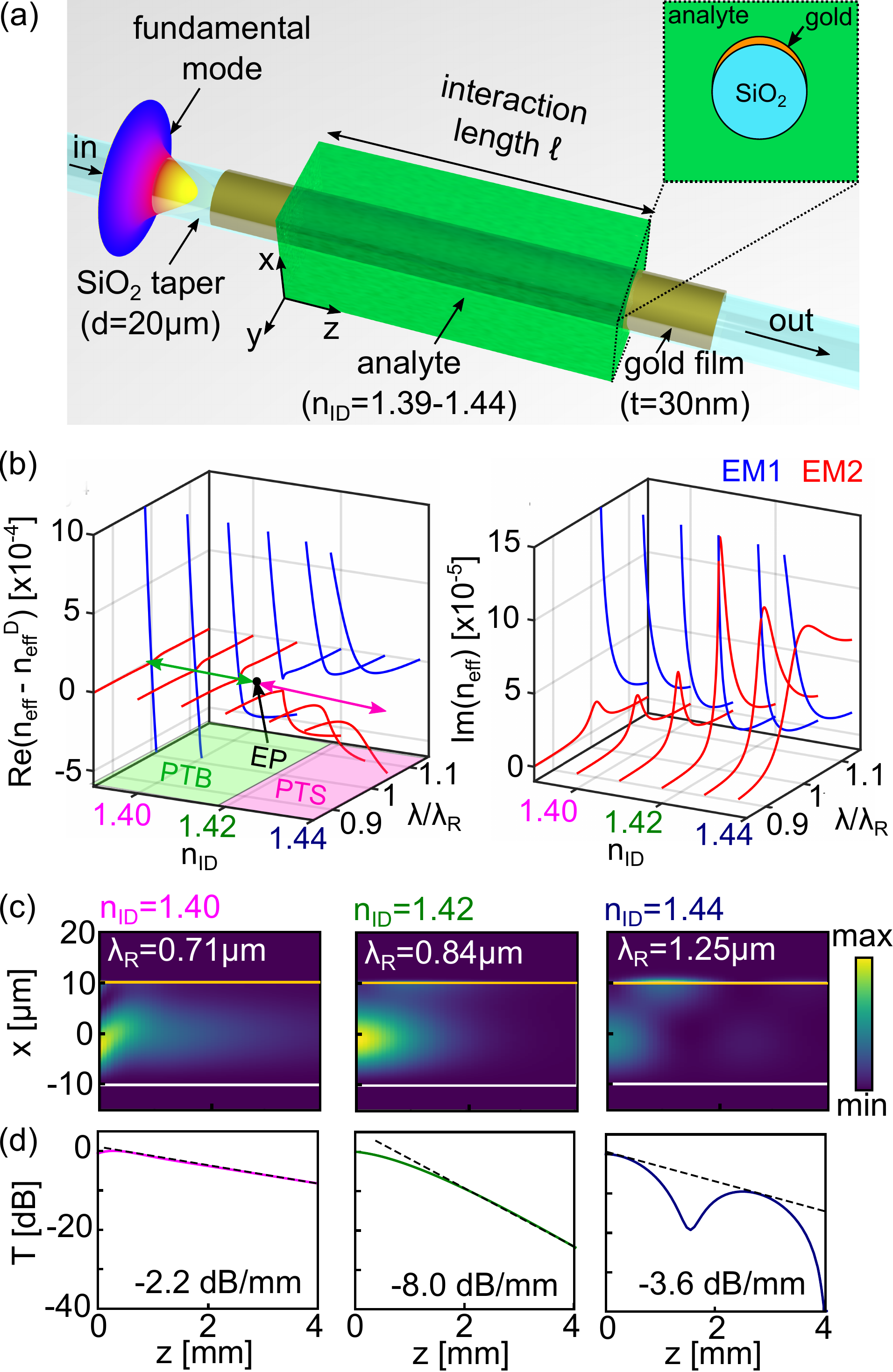}
\caption{\label{fig:fig2}(a) Schematic of the tuneable plasmonic/dielectric waveguide system considered. A cylindrical SiO$_2$ waveguide, coated by a gold nanofilm, is immersed in a liquid with predefined RI. (b) Spectral distribution of $\rm{Re}(n_{{\rm eff},j}-n_{\rm eff}^D)$ ($j=1,2$; $n_{\rm eff}^D$: effective index of equivalent dielectric waveguide without gold film) and ${\rm Im}(n_{\rm eff}$) for EM1 and EM2 as function of $\lambda/\lambda_R$ (see SM for $\lambda_R$ values.) (c) Spatial distribution of the Poynting vector and (d) transmitted power into the dielectric waveguide at three example RIs. Dashed lines: fitted loss as labelled.}
\end{figure}

To experimentally reveal the phase transition between PTB and PTS regimes across the EP, we consider a dielectric/plasmonic hybrid waveguide system with properties that can be tailored via the environmental refractive index (RI) (Fig.\,\ref{fig:fig2}(a)). The system is formed by a cylindrical SiO$_2$ waveguide (diameter: $d = 20\,{\mu {\rm m}}$) coated with a gold nano-film on one side (thickness: 30\,nm). The EM dispersions are tuned by immersing the waveguide in liquids with different RI, with the interaction length determined by the length of the liquid column $\ell$~\cite{wieduwilt2015ultrathin}. The uncoupled modes satisfy the phase matching condition at a wavelength $\lambda_R$ only when the outer RI has a specific value -- a scheme which has frequently been used for plasmonic sensing\,\cite{nov,wieduwilt2015ultrathin}.  Here, we use commercially available analytes, identified by the refractive index $n_{\rm ID}$ at $\lambda=589\,{\rm nm}$ (see Supplemental Material (SM) for the measured analyte and gold RI dispersions $n_o$).

We now consider the Eigenstates of this system when changing the RI environment, and how this impacts coupling. Since $d\gg\lambda$, a qualitative understanding of the modal behaviour can be obtained using a 1D multilayer system that includes the RI distribution along the connection line between the centre of the silica core and the azimuthal location of maximum gold film thickness~\cite{wieduwilt2015ultrathin}. The evolution of the Eigenstate dispersions for this hybrid system (Fig.\,\ref{fig:fig2}(b), where $n_{{\rm eff},j} = \beta_j^{\rm EM}/k_0$, $k_0 = 2\pi/\lambda$) show a clear transition from PTB to PTS regimes. The EP is predicted close to $n_{\rm ID}=1.42$, but its exact location will be highly sensitive to environmental conditions~\cite{chen2017exceptional}. Note that the detailed properties of hybrid multimode dielectric/plasmonic waveguides are more complex than what the CMT model predicts, also because SPP modes are close to cut-off, with obvious implications on modal dispersion.

Calculations show that the transition across the plasmonic EP can be inferred from the transmission properties of this waveguide system by using different $\ell$ and RI environments. To illustrate this, we calculate the axial intensity distribution (Fig.~\ref{fig:fig2}(c)), and retrieve the transmitted power (Fig.~\ref{fig:fig2}(d)) using the modal evolution and interference of the two hybrid EMs excited by the fundamental mode of the corresponding dielectric waveguide~\cite{tuniz2016broadband}. The complex modal amplitudes at input and output are determined by overlap integrals between the fields at $\ell$ and the fundamental mode of the dielectric waveguide, allowing to calculate the transmission through the hybrid system. The plots in Fig.~\ref{fig:fig2}(c) are calculated at $\lambda_R$, which in the PTB or PTS regime respectively occurs where the imaginary- or real- parts of $n_{\rm eff}$ cross (see SM) As expected, the PTS regime ($n_{\rm ID}=1.440$) shows an oscillating damped energy distribution between dielectric core and plasmonic layer (Fig.~\ref{fig:fig2}(c), right), due to modal beating of lossy EMs (Fig.~\ref{fig:fig2}(d), right). In the PTB regime ($n_{\rm ID}=1.400$), no such oscillations are observed (Fig.~\ref{fig:fig2}(c), left), and the transmitted power decays monotonically at smaller loss values (Fig.~\ref{fig:fig2}(d), left). For $n_{\rm ID}=1.420$ we find an intermediate case, where the transmission monotonically decays with the higest overall loss. Therefore, the two regimes can be distinguished by the axial power distribution, with the PTS case yielding an oscillatory behaviour, while the PTB case shows a monotonic power decay, both of which have lower overall loss with respect to the loss at the EP (dashed lines). The fact that the EP possesses the highest overall loss in this kind of hybrid plasmonic system has been verified by a detailed numerical analysis (see SM) Since the loss of this system is maximum at the EP, the presence of a local loss maximum provides an additional property that can be used to confirm that a $\mathcal{PT}$-phase transition through the EP has occurred.
Note that the simulations presented (Fig.~\ref{fig:fig2}(d)) enable a qualitative comparison to the experimental data and have been conducted in order to justify the choice of the hybrid waveguide used; the strong susceptibility on environmental influences, however, makes a quantitative comparison to experiments challenging.

\begin{figure}[t!]
\centering
\includegraphics[width=0.8\linewidth]{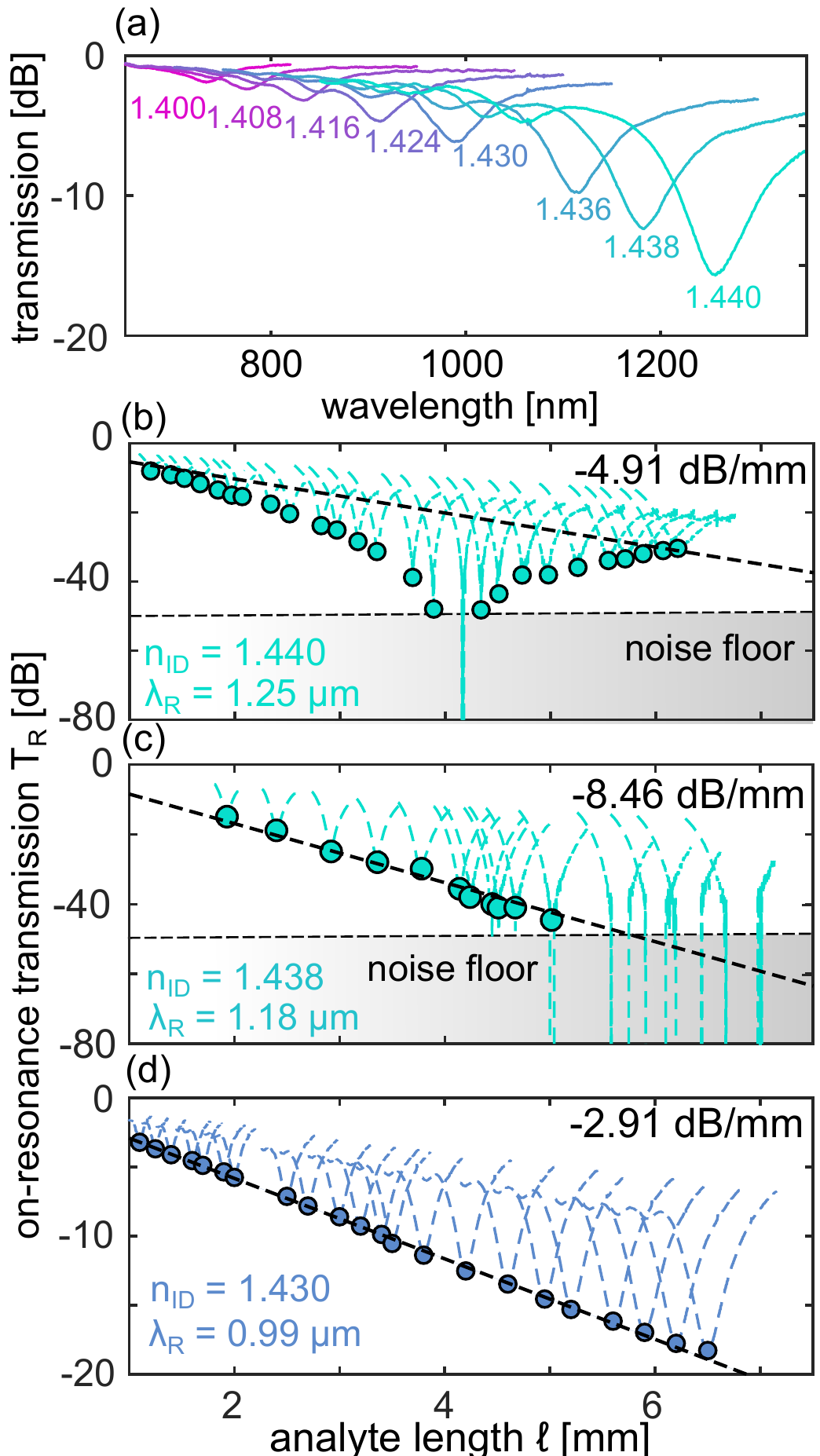}
\caption{\label{fig:fig4} (a) Measured spectral distribution of the power transmission through the hybrid waveguide (Fig.~\ref{fig:fig2}, $\ell = 1\,{\rm mm}$). Circles: measured on-resonance transmission as function of $\ell$ for the analytes as labelled, showing the transition from (b) the PTS to (d) the
PTB regime via (c) high-loss modes close to the EP. For each transmission dip, we have included the full resonance spectrum centered in each transmission minimum (i.e., the horizontal axis of each spectrum is scaled by a factor $\ell/\lambda_R$). Black dashed lines: fitted loss as labelled.}
\end{figure}

We perform a comprehensive series of transmission experiments on the above system -- details of sample preparation and experimental setup can be found in Ref.~[21]. Here, different interaction lengths were implemented using Teflon stubs of different length, which support defined droplets of analyte (spatial resolution: $\pm$ 0.05\,mm.)
We first measured the spectral distribution of the transmission $T$ for various analytes at constant interaction length ($\ell = 2\,{\rm mm}$, Fig.~\ref{fig:fig4}(a)). We observe a transmission dip that shifts towards longer wavelengths and increases in contrast as $n_{\rm ID}$ is increased. The minimum transmission ($T_{\rm R}$) for each $n_{\rm ID}$ occurs at resonant wavelengths $\lambda_R$. Calculations show that the transmission minimum is due to the interactions between the input mode and the excited hybrid EMs excited (see SM) and depends on both $\ell$ and the effective $\mathcal{PT}$-phase accessed. By analysing $\lambda_R$ as function of RI at $\lambda_R$, we determined the experimental RI sensitivity given by $S=\partial\lambda_R/\partial n_o$, which exceeds $50\rm \mu m/RIU$ for $n_o>1.43$ (see SM) This represents one of the highest values ever measured for any plasmonics device to date, and is due to operation near the cutoff~\cite{wu2009ultrasensitive} of the plasmonic mode.

Based on our previous analysis, we identify the transition between the two regimes by the different spatial power distribution~(Figs. \ref{fig:fig2}) by measuring the on-resonance power $T_R$ vs. $\ell$ at $\lambda_R$ under three different conditions. A high analyte RI ($n_{\rm ID}=1.440$, Fig.~\ref{fig:fig4}(b)) shows a strong dip in the spatial power transmission, indicating periodic energy exchange between the waveguides, i.e., directional coupling, confirming that this configuration is in the PTS regime.
In contrast, a lower analyte RI (here $n_{\rm ID}\leq1.438$, Fig.\,\ref{fig:fig4}(c),(d)), yield an exponentially decreasing spatial power distribution with no dips, indicating no directional coupling, i.e., the PTB regime.

Note that due to power oscillations in the PTS case (Fig.\,\ref{fig:fig4}(b)), it is difficult to estimate the overall loss; however, from peak of the envelope we can estimate a lower bound of $\sim 4.91\,{\rm dB/mm}$.
This is higher than what is measured in the PTB regime (Fig.\,\ref{fig:fig4}(d), 2.91\,dB/mm), but lower than what is measured close to the EP (Fig.\,\ref{fig:fig4}(c), 8.46 dB/mm), in qualitative agreement with the model (Fig.\,\ref{fig:fig2}(b),(c)) --  a further confirmation that the EP was crossed.

This study reveals that the effective $\mathcal{PT}$-symmetric and $\mathcal{PT}$-broken phases can be tuned through the EP in a non-Hermitian plasmonic system by adjusting the environmental refractive index. The PTB regime is particularly suitable for bioanalytic sensing, as the power transmission only weakly depends on propagation distance at moderate loss level. This allows to probe thin layers with large lateral dimensions (e.g., molecular layers or 2D materials), since the evanescent field is homogeneously probing the material of interest along the entire waveguide, in contrast to all-dielectric systems. The PTS regime instead allows efficient and rapid energy transfer into plasmonic layers, leading to the extraordinary high RI sensitivity measured, which can be employed for highly-demanding RI sensing applications, e.g., detecting single DNA-binding events or protein folding.

We have found that modes at the EP possess the highest overall loss for this particular system, a property which could be harnessed for plasmonic sensing applications near the EP. A recent theoretical report presented a similar result in a different context~\cite{ke2018strong}, warranting further study to verify its generality. Additionally, the phase indices of the EMs show a strong dependence of the effective mode index on the environmental RI, with additional implications for sensing. A modification of our structure could yield passive devices with extreme sensitivity  at desired RIs (e.g., at the water index) to detect individual molecular events~\cite{chemnitz2016enhanced}.

Our platform extends the capabilities of tuneable non-Hermitian photonics near the EP~\cite{miri2019exceptional}. The results and concepts presented here apply to any non-Hermitian waveguide, and immediately provide design tools for future EP-based photonic devices. Due to the unique properties of the two regimes and the unique dispersion characteristics near the EP, our study widens the scope of applications areas of plasmonics, with applications in bioanalytics, nonlinear light generation, signal processing, quantum technologies, and topological physics.

A.T. acknowledges support from the University of Sydney Postdoctoral Fellowship scheme.

\bibliography{main}

\clearpage

\onecolumngrid

\renewcommand{\thefigure}{S\arabic{figure}}
\setcounter{figure}{0}

\section{Supplemental Material}

\section{Coupled Mode Theory}

In order to gain insight into the modal behaviour of hybrid plasmonic/dielectric waveguides (Fig. 1(a) in the manuscript), we develop a coupled mode theory (CMT) approach that allows us to understand  Eigenmode (EM) formation inside the unbalanced two-waveguide systems considered. Our waveguides support one lossless (dielectric-like) mode ($\beta_1=\beta_1^R$) and one lossy (plasmonic-like) mode with strong modal attenuation and complex wave vector ($\beta_2=\beta_2^R+i\beta_2^I$)
where $\beta_2^I$ is assumed constant. The EMs  of the coupled mode system are described by superpositions of the isolated waveguide states, which holds true when the modes are moderately coupled~\cite{yariv2006photonics}. To include waveguide dispersion into the analysis, we assume that the phase indices $\beta_j^R$ of the waveguide modes have a linear dependence on frequency within a spectral region around a resonance frequency $\omega_R$:
\begin{equation}
\beta_{j} ^R= \beta_{j}^{R}(\omega_R) + \frac{\Delta\omega}{{v_{j}^G\vert_{\omega_R}}},
\label{eq:Taylor}
\end{equation}
where $v_j^G$ is the group velocity and $\Delta\omega = \omega - \omega_R$ is the frequency detuning. For simplicity, we assume that  $\omega_R$ is at the phase-matching frequency of the isolated modes, $\beta_1^{R}(\omega_R) = \beta_2^{R}(\omega_R)$. A relevant parameter is the mode dephasing given by $\Delta\beta_R = (\beta_1^R-\beta_2^R) /2= d \cdot \Delta\omega$, which is zero at $\omega_R$ and includes the group velocity mismatch  $d=(v_{(1,G)}^{-1} -v_{(2,G)}^{-1})/2$. According to the procedure to find Eq.~(2) in the main manuscript, the dispersion of the two hybrid Eigenmodes of the system is given by:
\begin{equation}
\beta_{j}^{\rm EM} = \overline{\beta_R}+i\beta_2^I/2\pm\sqrt{(d \cdot \Delta\omega + i\beta_2^{I}/2)^2 +\kappa^2},\label{eq:DispEig}
\end{equation}
where $\kappa$ is the coupling constant (assumed to be real-valued), and $\overline{\beta_R}$ is the average wave vector. At the phase-matching point of the isolated modes, $\Delta\omega=0$ and the dispersion of the EMs is given by  $\beta_j^{EM} = \overline{\beta_R} + \frac{i}{2}\beta_2^I \pm \Delta\beta^{EM}$, where $\Delta\beta^{EM} =\sqrt{\kappa^2 - (\beta_2^I/2)^2}$ is the resonance EM dephasing parameter. This parameter is essential for the subsequent discussion since the square root term indicates that the behaviour of the EMs is governed by the sign of the argument under the square root. A splitting with respect to the phase index between the EMs is obtained in case the argument is positive ($\kappa<\beta_2^I$), corresponding to the  $\mathcal{PT}$-symmetric situation. A negative argument ($\kappa>\beta_2^I$) leads to a complex square root, which imposes a splitting in the dispersions of the imaginary part of the wave vectors,  representing the $\mathcal{PT}$-broken situation. Note that if both waveguides are lossless, $\beta_2^I < \kappa$ and thus energy exchange is provided for any value of coupling constant.

The exceptional point occurs when the complex EM propagation constants coalesce ($\beta_1^{EM}=\beta_2^{EM}$). For real $\kappa$, the exceptional point occurs at the phase-matching point of the isolated modes (at $\Delta\omega=0$), leading to the \emph{exceptional point condition} (EPC) $\kappa = \beta_2^I/2$, as discussed in the manuscript. Note that the EPC is solely defined by the coupling of the waveguides and the loss of one mode.

For the example shown in first figure, we assume $\beta_j^R(\omega_R)=n_jk_0$ and $v_j^G=c_0/N_j$ ($c_0$: speed of light). Here $n_j$ and $N_j$ are the phase and group indices of the associated modes at $\omega_R$. The imaginary part of the wave vector is defined by $n_2^I=\beta_2^I/(\omega_R/c_0)$. The following parameters have been considered, which are not related to the actual experiments but are rather chosen to  emphasize the difference between the two $\mathcal{PT}$ regimes (Fig.~S1(a)).

\begin{table}[h!]
\begin{tabular}{|c|c|c|c|c|}
\hline
mode & $n_j$ & $N_j$ & $n_2^I$    \\
\hline
1    & 1.4  & 1.0   & 0       \\
\hline
2    & 1.4   & 3.4   & 0.127      \\
\hline
\end{tabular}
\caption{Modal parameters used to calculate the modal dispersions shown in Figs. 1(a) and S1(a).}.
\end{table}

\newpage

\section{Comparison between simulations}

\begin{figure*}[h!]
\centering
\makebox[\textwidth][c]{\includegraphics[width=1.0\textwidth]{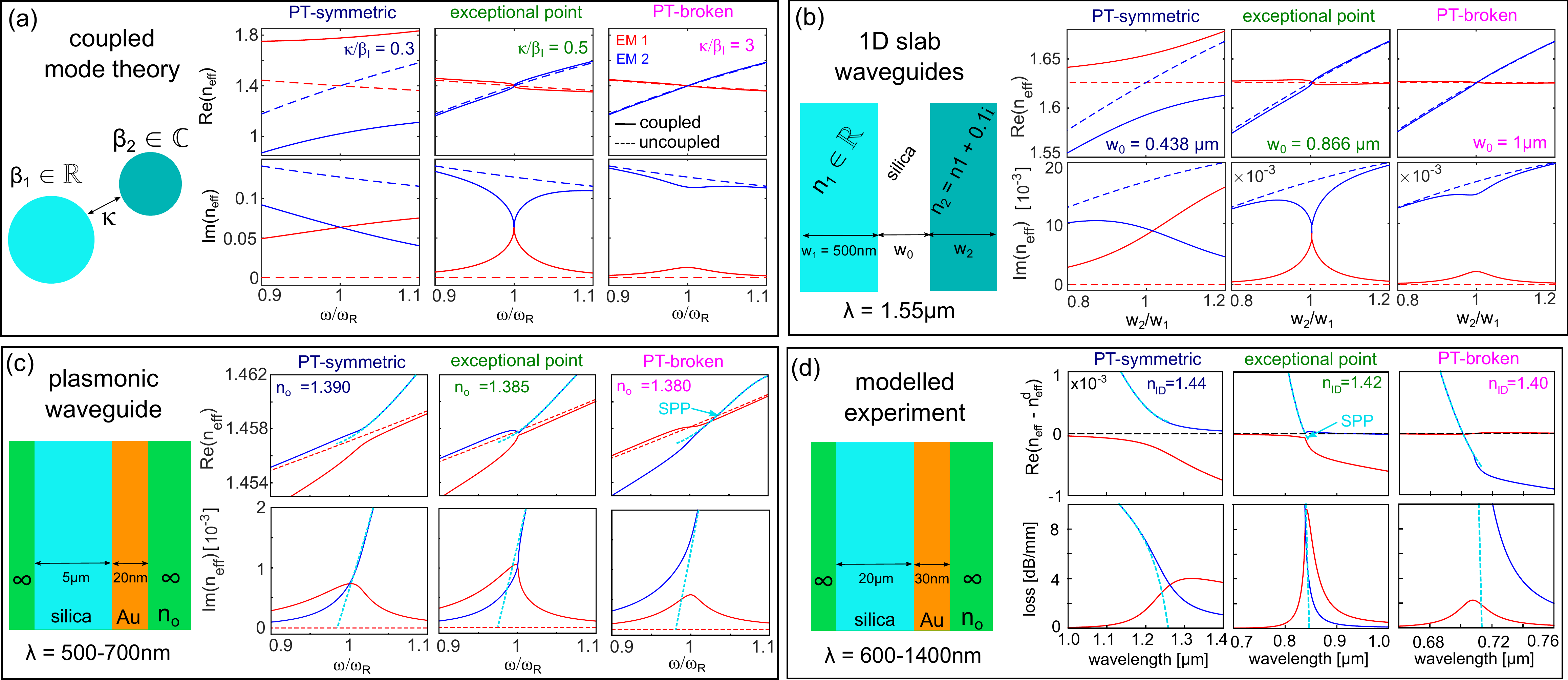}}%
\caption{Comparison between (a) coupled mode theory, (b) numerical solution of 1D slab waveguide problem, (c) an example plasmonic system, (d) and our modelled experiment (dark blue: PTS case (dark blue), green: EP, magenta: PTB case). To simplify comparison between the theoretical considerations, (a)-(c) plot the real and imaginary parts of the EM effective indices ($n_{{\rm eff},i} = \beta_i/k_0$, $k_0 = 2\pi/\lambda$) vs. normalized parameters (main text for details). For ease of comparison with Fig.~2 in the main manuscript, (d) shows ${\rm{Re}}(n_{{\rm eff},i}-n_{\rm eff}^d)$ ($i=1,2$) ($n_{\rm eff}^d$: effective index of equivalent dielectric waveguide without gold film) and loss (in dB/mm) for EM1 and EM2 as function of wavelength. A schematic for each geometry, including the material distribution and the parameters used, has been included in each subfigure. Blue and red curves represent the hybrid EM1 and EM2 modes, respectively. The light blue dashed lines correspond to the situation of an equivalent surface plasmon polariton in the absence of a dielectric waveguide and which is therefore not hybridized, i.e., forms inside a silica/gold/analyte system. See text below for further details of the models and situations represented.}
\end{figure*}

\subsection{Multilayer Model - Lossy Slab Waveguides}

Figure~S1(b) shows the real and imaginary parts of the effective index obtained by numerically solving Maxwell's equations to obtain the bound Eigenmodes of a 1D multilayer slab waveguide in TM polarization using appropriate boundary conditions~\cite{li1987general, tuniz2016broadband}. As an example, we take parameters similar to those used in one of our earlier works~\cite{tuniz2016broadband}, and consider a lossless waveguide (WG1) with refractive index $n_1 = 1.87$ and fixed width $w_1= 500,{\rm nm}$ embedded in a silica background~\cite{malitson1965interspecimen}, adjacent to a lossy waveguide (WG2) with complex refractive index $n_2 = 1.87 + 0.027i$. Phase matching occurs when $w_1 = w_2$. To simplify calculations, we consider a fixed wavelength $\lambda = 1.55\,\mu{\rm m}$ and vary the width $w_2$ of WG2 for different edge-to-edge separations $w_0$, which in the absence of material dispersion is equivalent to changing $\lambda$ for constant geometry. The resulting dispersions (Fig.~S1(b)) clearly show that the three regimes shown in Fig.~S2(a) can be accessed by changing the edge-to-edge separation between the two waveguides, which overall is equivalent to changing the coupling constant. The power exchange characteristics between the two waveguides (i.e., the longitudinal power distribution) in these three regimes are shown in Figs.~1(b) and 1(c) of the main manuscript.

\subsection{Multilayer Model - Example Hybrid Dielectric-Plasmonic Waveguide}

To minimize calculations time ($\sim 3\,{\rm ms}$ per wavelength) and elucidate the generality of the underlying physics for the experimentally measured hybrid dielectric-plasmonic system, we consider a smaller version of the device  used in the experiment as quantitative example. We again consider a 1D multilayer system, here formed by a silica (${\rm SiO}_2$) slab waveguide~\cite{malitson1965interspecimen} (slab width: $d=5\,\mu{\rm m}$) in contact with a gold nanofilm ($t=20\,{\rm nm}$). The entire structure is surrounded by a material with constant refractive index $n_o = n_{\rm ID}$. As per our experiment, changing the surrounding refractive index changes the coupling condition between the dielectric and plasmonic waveguide, revealing the three different regimes. For ease of comparison with Fig.~S1(a),S1(b), Fig.~S1(c) shows the real and imaginary parts of the effective indices obtained as a function of the normalized frequency ($\omega_R = 3.30 \times 10^{15}\,{\rm Hz}, 3.26 \times 10^{15}\,{\rm Hz}, 3.23 \times 10^{15}\,{\rm Hz}$ for $n_{\rm ID} = 1.380, 1.385, 1.390$, respectively), once again illustrating that the three regimes of Fig.~S1(a) can be accessed. This geometry will be used to highlight some of the most important properties of this system --  see ``Crossing the EP in the hybrid plasmonic-dielectric waveguide''.

\subsection{Multilayer Model - Experimental Hybrid Dielectric-Plasmonic Waveguide}

Finally, we model the experimental case discussed in the main manuscript by approximating it as a slab waveguide. The most important differences with respect to the previous case are different geometric parameters ($d=20\,\mu{\rm m}$ and $t=30\,{\rm nm}$), and that the experimentally measured refractive index of the analytes are included (Fig.~S3(a)). We also include the experimentally measured dispersion of gold (Fig.~S3(b)) and silica~\cite{malitson1965interspecimen}. Note that the large waveguide extension demands a high numerical-precision package to solve the boundary problem, and the solver is $~100\times$ slower than the previous case. The spectral distributions of phase index and modal loss are shown in Fig.~S1(d), for  comparison with the experiment shown in the main manuscript (Fig.~2(b)). Similarly to the previous case, we notice that for the analyte indices considered, we transition between the $\mathcal{PT}$-broken (PTB) and $\mathcal{PT}$-symmetric (PTS) regimes via the EP.

\begin{figure}[t!]
\includegraphics[width=0.75\textwidth]{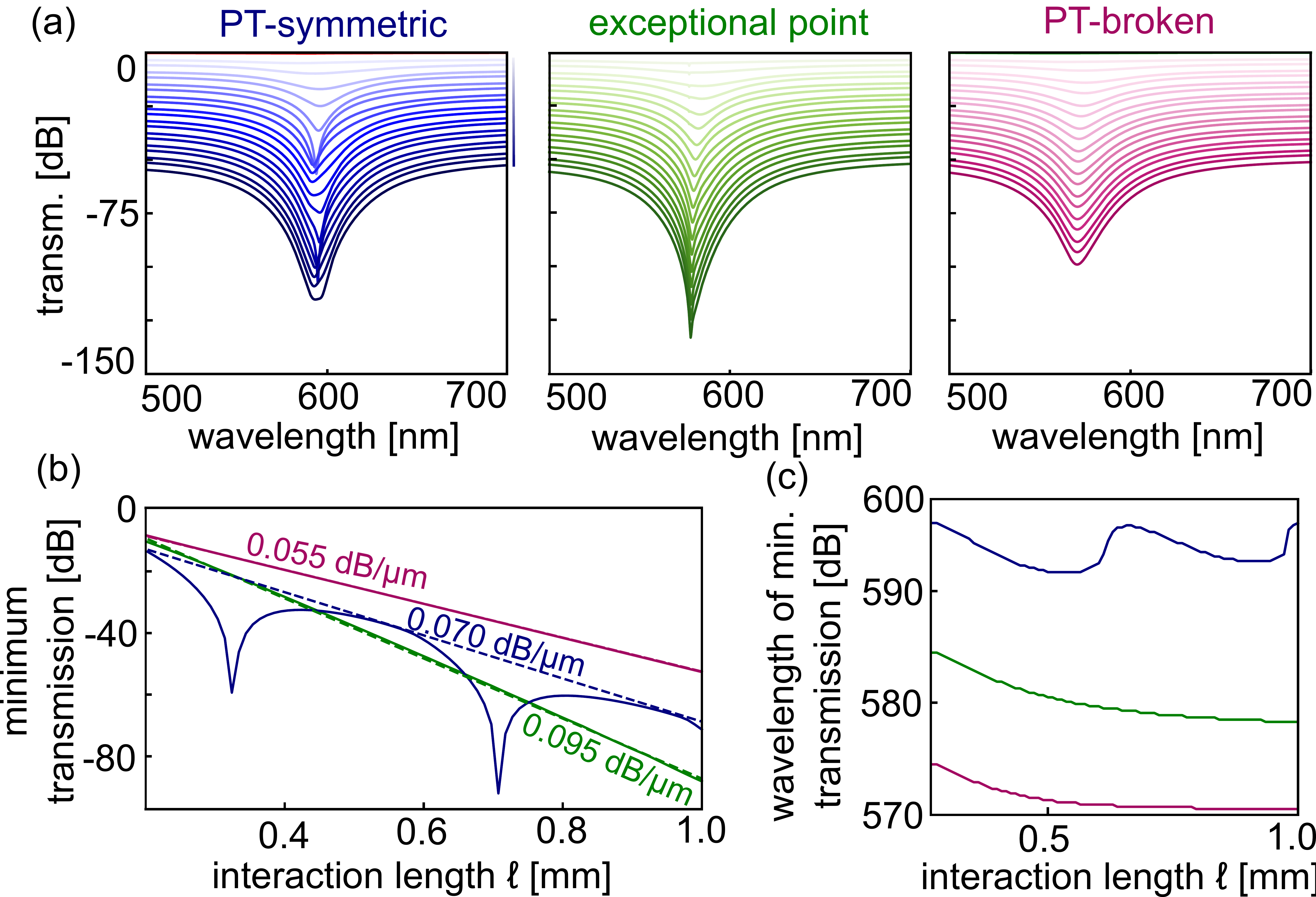}%
\caption{(a) Transmission spectra for the three cases of Fig. S1(c) for increasing analyte length ($\ell = 0.1-2\,{\rm mm}$, light to dark). (b) Minimum transmission as function of $\ell$ for the three different regimes shown in (a). Note that the loss at the EP is the largest. (c) Corresponding wavelengths of minimum transmission as function of $\ell$.}
\end{figure}

\section{Crossing the EP in the hybrid plasmonic-dielectric waveguide}

Using the method presented in Ref.~\cite{tuniz2016broadband}, we calculate the spectral distribution of the transmission of the example hybrid plasmonic waveguide for different propagation lengths $\ell$, which in the experiment are defined by the length of the liquid column. Specifically, we obtain the power in the dielectric waveguide by evolving the hybrid Eigenmodes, each with its respective complex propagation constant over the distance $\ell$. The contribution of each EM at input is obtained from the overlap integral between the fundamental mode of the dielectric waveguide and the two plasmonic hybrid EMs. The total output field is projected onto the dielectric mode at output, with the modulus squared of the resulting amplitude yielding the output power. This procedure provides a rapid and efficient method to calculate the spectral distribution of the transmission as a function of $\ell$, for each regime presented in Fig.~S2(c), while accounting for: (i) the $\mathcal{PT}$-phase dependent excitation of each hybrid dielectric-plasmonic Eigenmodes by the fundamental dielectric mode at input; (ii) evolution up to end of the coupling section of the hybrid waveguide system.

Figure~S2(a) shows a waterfall plot of the transmission as a function of wavelength for the three regimes of Fig.~S1(c), for the interaction length domain $0.1<\ell< 2\,{\rm mm}$. Each transmission plot has been offset by 2\,dB for increasing values of $\ell$ for clarity. We observe that each regime possesses distinct spectral features, which can be conveniently discerned from their properties at the point of minimum transmission. Figure~S2(b) shows the minimum transmission  as a function of the propagation length, and Fig.~S2(c) shows the wavelength at which the minimum transmission occurs as a function of $\ell$. We find that the hybrid dielectric-plasmonic design enables the identification of the exceptional point via two independently measurably characteristics while transitioning from the $\mathcal{PT}$-symmetric to the $\mathcal{PT}$-broken regime: (1) the transmission of the dielectric waveguide at resonance transitions from an oscillating decay (periodic energy transfer between waveguides, (Fig.~S2(c), blue), to exponential decay (no energy transfer between waveguides, Fig.~S2(c), purple) (2) coupling between waveguides at the exceptional point leads overall to the largest attenuation (Fig.~S2(b), green).  The EM characteristics of this 1D system (Fig.~S1(c)) is qualitatively similar to that of our modelled experiment (Fig.~S1(d) and Fig. 2 of the main manuscript), which in turn has been confirmed by experiment (Fig.~3 in the main text).

Note that in the PTS regime (regime where the imaginary parts of $n_{\rm eff}$ show an avoided crossing), we find that the overall loss at resonance has the same value as to the maximum loss of the (low-loss) EM1.
In the PTS regime, the overall loss of the system is the envelope of the periodically oscillating transmission, because power is being transferred between waveguides: the fitted loss value obtained from the envelope (Fig. S2(b), blue) coincides with the loss at the wavelength where the imaginary parts of the two EMs intersect (Fig. S1(c)).

\section{Ellipsometry measurements of analytes and gold film}

\begin{figure}[h!]
\includegraphics[width=\textwidth]{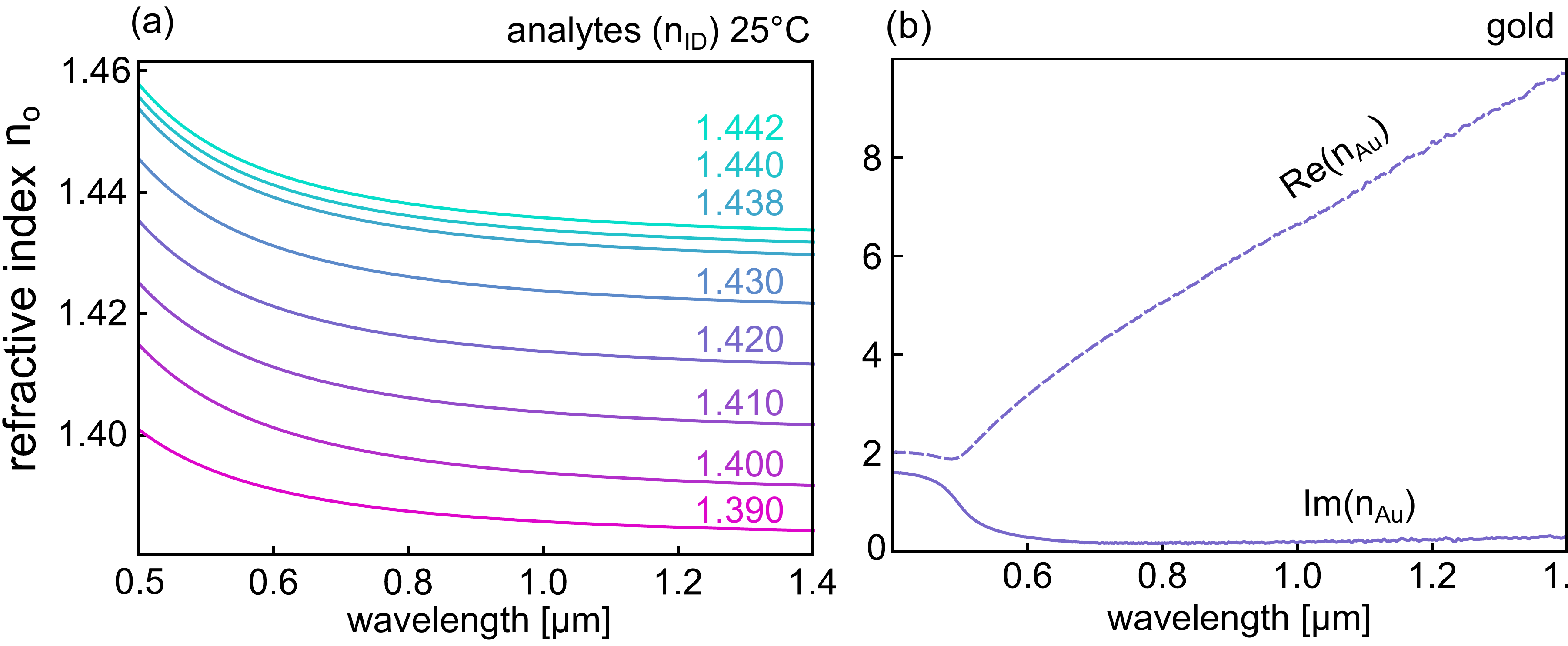}%
\caption{Measured spectral distribution of the refractive indices $n_o$ of (a) the analytes (identified by the label $n_{\rm ID}$) and (b) a 30\,nm gold film both determined using a variable-angle spectroscopic ellipsometer (Sentech, SE850).}
\end{figure}

\newpage

\section{Sensitivity}

The RI sensitivity ($S = \Delta \lambda_R/\Delta n_o$) was calculated from the measured resonance wavelengths $\lambda_R$ as a function of measured analyte RI $n_o$ for a fixed column length of 2\,mm, using a previously reported approach~\cite{wieduwilt2015ultrathin}. The resonance wavelengths were determined using a second-order polynomial fitting applied to the minima of the normalized transmission spectra. The numerical derivative of the resulting curve with respect to $n_a$ results in the sensitivity shown in Fig.~S4.

\begin{figure}[h!]
\includegraphics[width=0.6\textwidth]{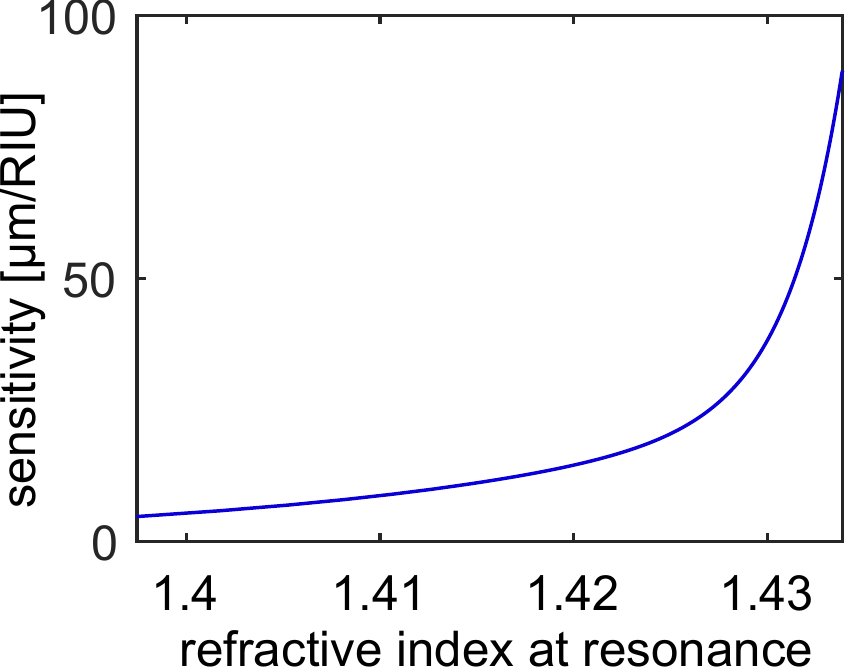}%
\caption{Refractive index sensitivity as function of the refractive index at resonance, determined by differentiating a second-order polynomial to the experimentally obtained relation $\lambda_R=\lambda_R(n_o)$.}
\end{figure}

\end{document}